\documentclass[aip,graphicx]{revtex4-1}

\usepackage{graphicx}
\usepackage{amsmath}

\draft 

\begin{document}


\title{Nanoscale tip positioning with a multi-tip scanning tunneling microscope using topography images} 

\author{Arthur Leis}
\thanks{The authors to whom correspondence may be addressed: a.leis@fz-juelich.de, b.voigtlaender@fz-juelich.de.}
\affiliation{Peter Gr\"{u}nberg Institut (PGI-3), Forschungszentrum J\"{u}lich, 52425 J\"{u}lich, Germany}
\affiliation{J\"ulich Aachen Research Alliance (JARA), Fundamentals of Future Information Technology, 52425 J\"ulich, Germany}
\affiliation{Experimentalphysik IV A, RWTH Aachen University, Otto-Blumenthal-Stra\ss{}e, 52074 Aachen, Germany}

\author{Vasily Cherepanov}
\affiliation{Peter Gr\"{u}nberg Institut (PGI-3), Forschungszentrum J\"{u}lich, 52425 J\"{u}lich, Germany}
\affiliation{J\"ulich Aachen Research Alliance (JARA), Fundamentals of Future Information Technology, 52425 J\"ulich, Germany}

\author{Bert Voigtl\"ander}
\thanks{The authors to whom correspondence may be addressed: a.leis@fz-juelich.de, b.voigtlaender@fz-juelich.de.}
\affiliation{Peter Gr\"{u}nberg Institut (PGI-3), Forschungszentrum J\"{u}lich, 52425 J\"{u}lich, Germany}
\affiliation{J\"ulich Aachen Research Alliance (JARA), Fundamentals of Future Information Technology, 52425 J\"ulich, Germany}
\affiliation{Experimentalphysik IV A, RWTH Aachen University, Otto-Blumenthal-Stra\ss{}e, 52074 Aachen, Germany}

\author{F. Stefan Tautz}
\affiliation{Peter Gr\"{u}nberg Institut (PGI-3), Forschungszentrum J\"{u}lich, 52425 J\"{u}lich, Germany}
\affiliation{J\"ulich Aachen Research Alliance (JARA), Fundamentals of Future Information Technology, 52425 J\"ulich, Germany}
\affiliation{Experimentalphysik IV A, RWTH Aachen University, Otto-Blumenthal-Stra\ss{}e, 52074 Aachen, Germany}


\date{\today}

\begin{abstract}
Multi-tip scanning tunneling microscopy (STM) is a powerful method to perform charge transport measurements at the nanoscale. With four STM tips positioned on the surface of a sample, four-point resistance measurements can be performed in dedicated geometric configurations. Here, we present an alternative to the most often used scanning electron microscope (SEM) imaging to infer the corresponding tip positions.
After initial coarse positioning monitored by an optical microscope, STM scanning itself is used to determine the inter-tip distances. A large STM overview scan serves as a reference map. Recognition of the same topographic features in the reference map and in small scale images with the individual tips allows to identify the tip positions with an accuracy of about 20\,nm for a typical tip spacing of $\sim 1\,\mu$m. In order to correct for effects like the non-linearity of the deflection, creep and hysteresis of the piezo-electric elements of the STM, a careful calibration has to be performed.
\end{abstract}

\pacs{}

\maketitle 


\section*{Introduction}

The method of multi-tip STM has recently gained increased interest, as it has unique advantages when studying charge transport on the nanoscale. The standard approach for nanoscale charge transport measurements is to use lithographic methods for contacting nanostructures. An alternative approach for the contacting of nanostructures is to use the tips of a multi-tip scanning tunneling microscope, in analogy  to  the  test leads of a multimeter used at the macroscale.
The advantages of this approach are: 
(a) \textit{in situ} contacting of ``as grown'' nanostructures still under vacuum allows to keep delicate nanostructures free from  contamination which otherwise can be induced by lithography steps performed for contacting. 
(b) Flexible positioning of the contacting tips and different contact configurations are easy to realize, while lithographic contacts are fixed. 
(c) Probing with STM tips can be performed non-invasively (high ohmic), while lithographic contacts are typically invasive (low ohmic). 
Meanwhile, a fast growing community of researchers using multi-tip STM has evolved and it is already hardly possible to compile a comprehensive list of publications in this field. Here, we mention exemplary publications in this field \cite{Kanagawa,Fukui2020,Shimizu_ex1,Aono_ex2,Bobisch_ex1,TrompSTP,Polley,Hirose,Tegenkamp_ex1,Wolkow_ex1,Kolmer_ex1,Kolmer_ex2,Grandidier_ex1,Lord_ex1,Just1,Lupkecomm,Li_ex1,Civita2020,Lispin,Joachim_ex2,Gao_ex1,Kang}. Moreover, several reviews of the scientific work in this field are available \cite{Hofmann_rev,Aono_rev,Li_rev,Grandidier_rev,Miccoli,Aono_rev2,Voi_rev}. 

An important initial task in multi-tip scanning tunneling microscopy is to move the individual tips to the desired positions. This is done under observation of the tip positions with an scanning electron microscope integrated to the multi-tip STM instrument, or with an optical microscope monitoring the tips. As charge transport measurements on the nanometer scale are most desirable for a multi-tip STM, SEM imaging is preferred over the optical microscope due to its much higher spatial resolution. However, despite its much lower resolution, the use of an optical microscope also has advantages, as SEM imaging can lead to contaminations on the sample. In normal SEM imaging at pressures of $\sim 10^{-7}$\,mbar, the well-known carbon deposition on the sample due to decomposition of hydrocarbons in the residual gas occurs. However, it has also been shown that in some cases even the use of a UHV SEM, used at pressures of $\sim 10^{-10}$\,mbar can lead to SEM imaging induced contaminations on the sample \cite{Hallam2007} or to a change of the electrical properties of the sample \cite{Durand2013, Bloemers2012}. In this regard, the use of an optical microscope is beneficial, as it leaves the surface free of contaminants and does not alter the electrical properties of the sample. Moreover, also the cost and the effort to operate an SEM are much larger compared to an optical microscope.

Here we present a method to position all four STM tips in an area smaller than $1\,\mathrm{\mu m}^2$, with a precision of $\sim 20$\,nm for the determined inter-tip distances without the use of an SEM. Following an initial positioning of the tips under observation with an optical microscope with a resolution of about $2\,\mathrm{\mu m}$, imaging with the STM tips is used to move the four tips of the multi-tip STM to their desired positions. For this purpose, a large overview image (several $\mathrm{\mu m }$ width), obtained with one of the tips, is used as reference. Subsequently, small STM scans are recorded with each tip to identify topographic features that are also seen in the overview scan. Once a topographic feature is recognized, the position of the corresponding scanning tip is known. Further adjustment of each tip position can then be achieved by fine lateral movement of the tips. Potential sources of error in the tip positioning like thermal drift, piezo creep and the non-linearity of the piezo extensions have to be considered carefully in order to increase the position accuracy of the tips.

\section*{Tip positioning with the optical microscope}

\begin{figure}[t]
\includegraphics[width=\linewidth]{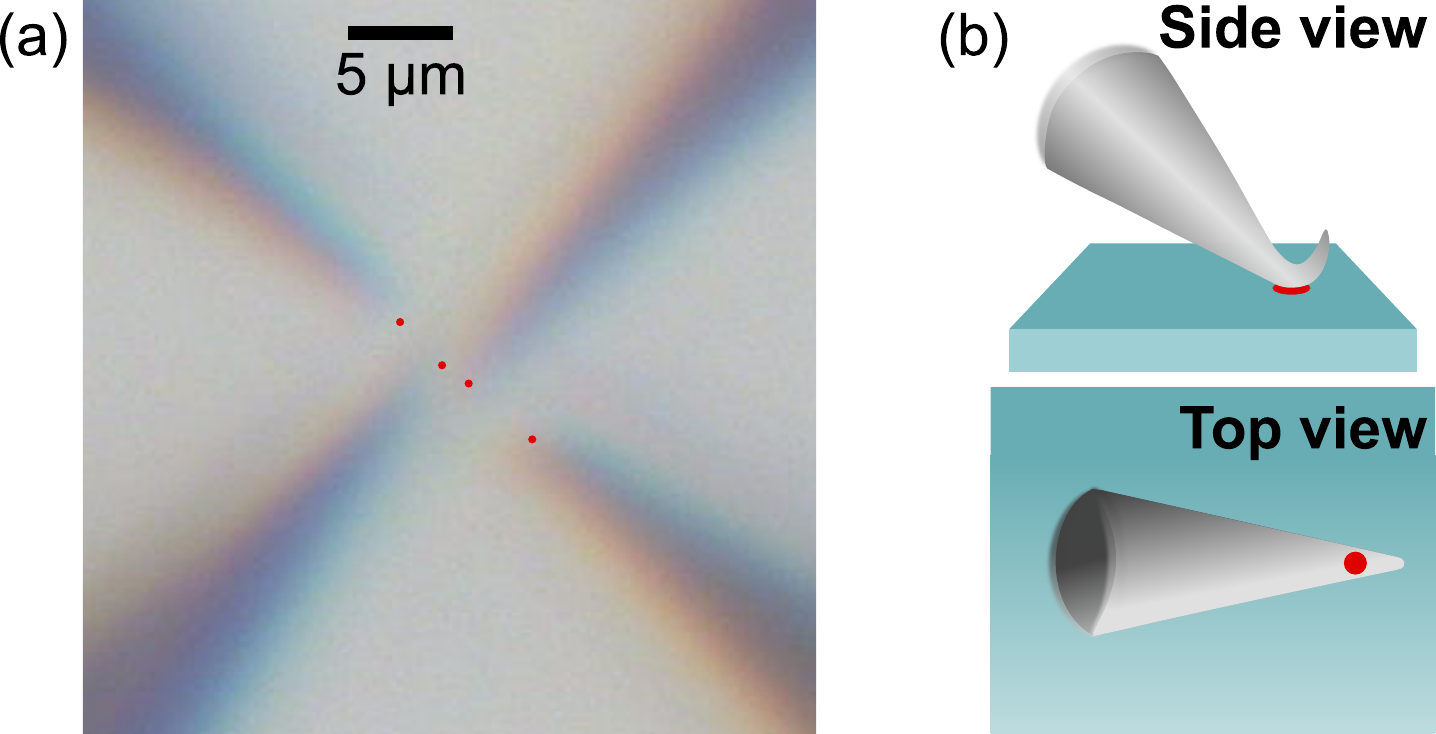}
\caption{
\textbf{(a)} Optical micrograph of an exemplary linear tip configuration used for four-point resistance measurements. When determining inter-tip distances from optical microscopy, the contact points of the tips are assumed to be at the end of the perceptible tip shape along the mirror symmetry axis, as indicated by the red dots. With inter-tip distances of $\sim 4\,\mu$m, the dimensions of the configuration are close to the resolution limit of the microscope. For smaller tip distances, the influence of the uncertainty is too large to reasonably determine the tip positions.
\textbf{(b)} Schematic illustration of a systematic position error due to a plastic tip deformation. When a tip is bent upwards, e.g. due to crude contacting, the respective deformation cannot be identified from the top view. As a result, the actual point of contact to a sample surface is shifted compared to its assumed position at the tip end as determined from the projection.}
\label{fig1}
\end{figure}

The experiments described here are performed with an ultra-high vacuum (UHV) multi-tip STM from mProbes \cite{Cherepanov2012}. The optical microscope used is a ZOOM 160 from EHD imaging GmbH and is directed towards the sample from below with a working distance of 50\,mm. A normal UHV NW40 viewport is used between the microscope (ambient) and the sample (UHV) \cite{Cherepanov2012}. 
An exemplary optical micrograph of a measurement configuration with inter-tip distances of $\sim 4\,\mu$m is shown in Fig.~\ref{fig1}~(a). When estimated from the top view provided by the micrograph, contact points of the tips are assumed to be at the perceptible end of each tip along the mirror symmetry axis of the tip, as denoted by the red dots in the image. 
In the presented example configuration, inter-tip distances are close to the microscope resolution, such that a significant uncertainty of the measured tip positions results. 
However, the precise knowledge of the tip positions on the sample and the tip-tip distances are required for nanoscale transport measurements. For instance, in a classical four-point measurement, the resistivity to be extracted depends directly on the tip-tip distances \cite{Voi_rev}. 

An additional problem in the determination of tip distances using an SEM or optical microscope is a possible introduction of systematic position errors due to tip deformations, as depicted in Fig.~\ref{fig1}~(b).
While lateral tip deformations, e.g. due to tip collisions, are easily seen from the top view of the tip arrangement, upward tip deformations cannot be identified, as the projected shape of the bent tip appears to be straight. Plastic upward deformations like this result from indenting a tip into hard surfaces. 
With tip positions nominally assumed to be at the end of the tip shape as it is perceptible in the microscope image, such a deformation introduces an additional error to the determination of inter-tip distances in the order of the bent length segment.
As this source of errors is due to the top view of the tip setup being a projection of three-dimensional objects onto the sample plane, the influence of this issue is also present when tip positions are monitored by SEM imaging.

\section*{Principle of tip positioning using STM scans}

In order to circumvent the outlined uncertainty issues and reach nanometer scale precision of tip positions, a positioning method relying on STM imaging is employed.
By using the topography of the sample surface as a basis for tip positioning, tips can be navigated across the surface in tunneling contact with greatly enhanced spatial precision.
Furthermore, by using STM scans for identifying desired tip placements on the sample topography, plastic tip deformations do not introduce positioning uncertainties, since the STM images are acquired in tunneling contact by the actual contact point itself. In principle, it can be difficult to obtain clear STM topography images with strongly deformed tips. In our experiments, the tips are indented into the surface with less than 10\,nm nominal depth, such that the damage is minimal. We generally find that it is still possible to acquire adequate scans even when the deformed tip section is up to $\sim 1 \, \mu$m in length.
Altogether, tip positioning via STM imaging enables a large range of possible tip configurations only limited by tip radii.

The general concept of the nanoscale positioning technique is to determine the position of each tip on the surface using the recorded topography of the underlying sample as a reference.
This approach is enabled by exploiting the ability of all four STM tips to function as individual scanners and is comprised of five steps:
\begin{itemize}
    \item[1.] Calibration of the non-linear piezo response of one of the four tips via optical microscope images.
    \item[2.] Recording of a large overview STM scan with the calibrated tip.
    \item[3.] Acquisition of small STM scans in the mapped overview area with all four tips to identify their current position via topographic features.
    \item[4.] Navigation of all four tips towards their target position in the overview scan.
    \item[5.] Lowering of all four tips to the sample surface for electric measurements.
\end{itemize}
As described in detail in the next section, step 1 is performed prior to the tip positioning by emulating the isolated $x$-motion of the scanning process of step 2. With step 2, the detailed topography of an investigated sample is related to spatial dimensions using the preceding calibration of the scanning tip. In principle, steps 1 and 2 can also be combined to perform a live calibration of the scanning tip displacement during the recording of the overview scan.
Steps 3 and 4 constitute the actual tip positioning and are performed iteratively until all tips are in the desired measurement configuration.

\begin{figure*}[t]
\includegraphics[width=\linewidth]{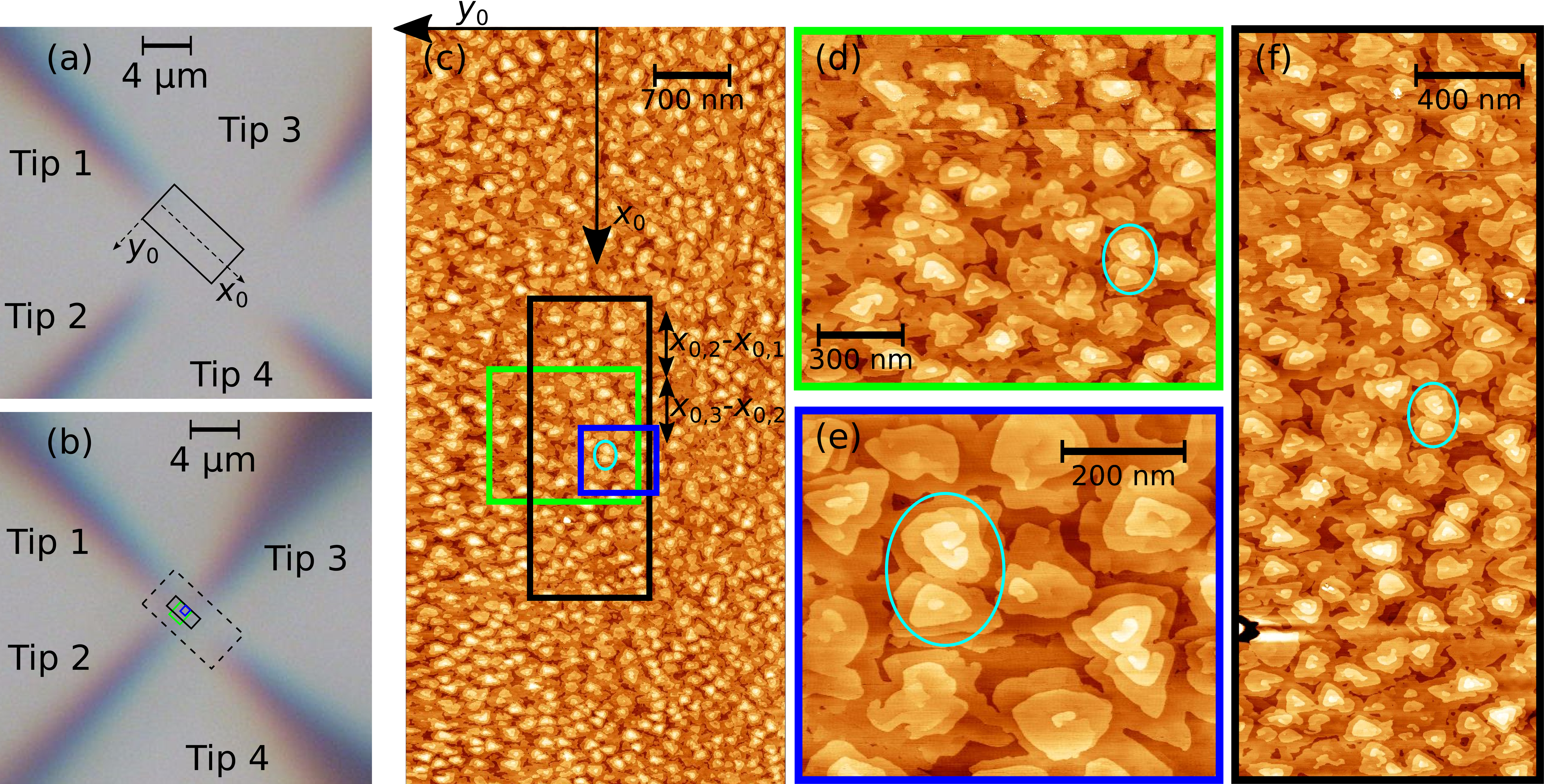}
\caption{Principle of the tip positioning procedure based on overlapping STM scans used for transport measurements on the nanoscale.
\textbf{(a)} Tip 1 is used to acquire a large overview scan, indicated as black rectangle. The recorded overview scan (shown in \textbf{(c)}) constitutes a reference map for precise tip navigation.
\textbf{(b)} Subsequently, all four tips are moved into the mapped area (monitored by the optical microscope) to perform individual small scans, which are shown by the three rectangles in \textbf{(c)}. STM images of the small scans are shown in \textbf{(d)-(f)}. Note that the presented example scans in \textbf{(d)-(f)} are not an enlarged view of the overview scan, but actual scans performed with tips 2, 3 and 4. The positions of all tips are identified within the reference map by recognition of topographic features from the corresponding small scans in the overview scan, as indicated exemplary by cyan ellipses for two particular islands.
After identifying all tip positions in the reference map, the tips are navigated to their target positions in the desired tip configuration. After every displacement step, the exact position of each tip can be reconfirmed by further scans.} 
\label{fig2}
\end{figure*}

The principle of steps 2 and 3 is illustrated in Fig.~\ref{fig2}, where a BiSbTe$_3$ sample was imaged \cite{Leis2020,Leis2021}. Prior to the scanning process of the overview scan, the other tips are moved out of the scan range to avoid collisions (Fig.~\ref{fig2}~(a)). 
One of the four tips (e.g. tip 1) is used to acquire an overview STM scan of a large area of the sample surface, as indicated in the optical microscope image Fig.~\ref{fig2}~(a) and presented as STM image in Fig.~\ref{fig2}~(c) for the case of a scan of $4\,\mu$m $\times 10\,\mu$m size. The scanned area is the area of interest for the eventually desired tip configuration (i.e. a linear tip configuration for a four-point measurement in this case). This STM topography scan serves as a reference map for the navigation of all tips. 

After the completion of the overview scan, the idle tips are moved into the scanned region, as indicated in Fig.~\ref{fig2}~(b). With the overview scan being large enough, the optical microscope offers sufficient accuracy to do so without tip collisions. Being situated in the scanned overview area, all tips are brought into tunneling contact to perform small scans of the sample surface. In case of the exemplary sample, these scans are presented in Fig.~\ref{fig2}~(d)-(f). It is noted that Fig.~\ref{fig2}~(d)-(f) present actual scans performed with tips 2, 3 and 4 and are not an enlarged view of Fig.~\ref{fig2}~(c). For the positioning procedure, it is crucial that all tips are located somewhere in the large overview scan in order to guarantee an overlap in the imaged topography with these small scans performed with each tip. 

By recognition of distinct features of the imaged sample topography, the small scans can be located in the image of the large overview scan (Fig.~\ref{fig2}~(c)). Prominent features that are useful for this purpose can be grainy step edges, islands or other characteristically shaped constituents of the topography. 
As an example, two characteristic islands (one with a screw dislocation) are highlighted by cyan ellipses in the four STM images shown in Figs.~\ref{fig2}~(c)-(f).
In our example system, the topography is rough and offers many characteristic features on the surface. If an investigated sample surface happens to have a plain, flat topography without distinct characteristics, still, steps and defects can be used for orientation in the scanned area. In more extreme cases of ideal samples with no defects, features can in principle be generated artificially on the surface prior to performing the overview scan. For this purpose, voltage pulses can be applied with the STM tips while in tunneling contact with controlled separations, leaving marks which can be used as anchor points.

With the small scans identified in the STM image of the large overview scan 
(Fig.~\ref{fig2}~(c)), the locations of the respective scanning tips within the overview scan are identified as well. In order to realize a desired tip arrangement on the mapped surface, the tips are navigated across the surface to their target position by fine $xy$-control of the scanning piezos without leaving tunneling contact. Fine movement in this fashion is controlled blindly by applying voltage to the scanning piezo without immediate visual confirmation. Therefore, tips are moved in steps with small STM scans performed in between to reaffirm the changed position within the overview image.

After navigating all tips to their target positions, all STM tips are lowered to the sample surface from tunneling contact to establish electrical contact \cite{Voi_rev} and perform the desired charge transport measurement, e.g. to perform a four-point measurement. If distance dependent four-point measurements are performed \cite{Just1,Just2017,Lupke2017,Leis2019,Leis2020}, the tip positions need to be varied in between electrical measurements. This is done by retracting the respective tips that are intended to be moved back into tunneling contact to navigate them to the next measurement position, while leaving the static tips at their position in hard contact.

From the realized tip configurations in a measurement series, the corresponding tip distances are determined from the positions of the tip in the large overview scan, as indicated in Fig.~\ref{fig2}~(c). By considering the non-linearity and the hysteresis of the displacement by the piezo element of the tip that is used for recording the overview scan, as shown in the following sections, distances in the scan can be determined more precisely.

\section*{Calibration of non-linear piezo response}

\begin{figure*}[t]
\includegraphics[width=\linewidth]{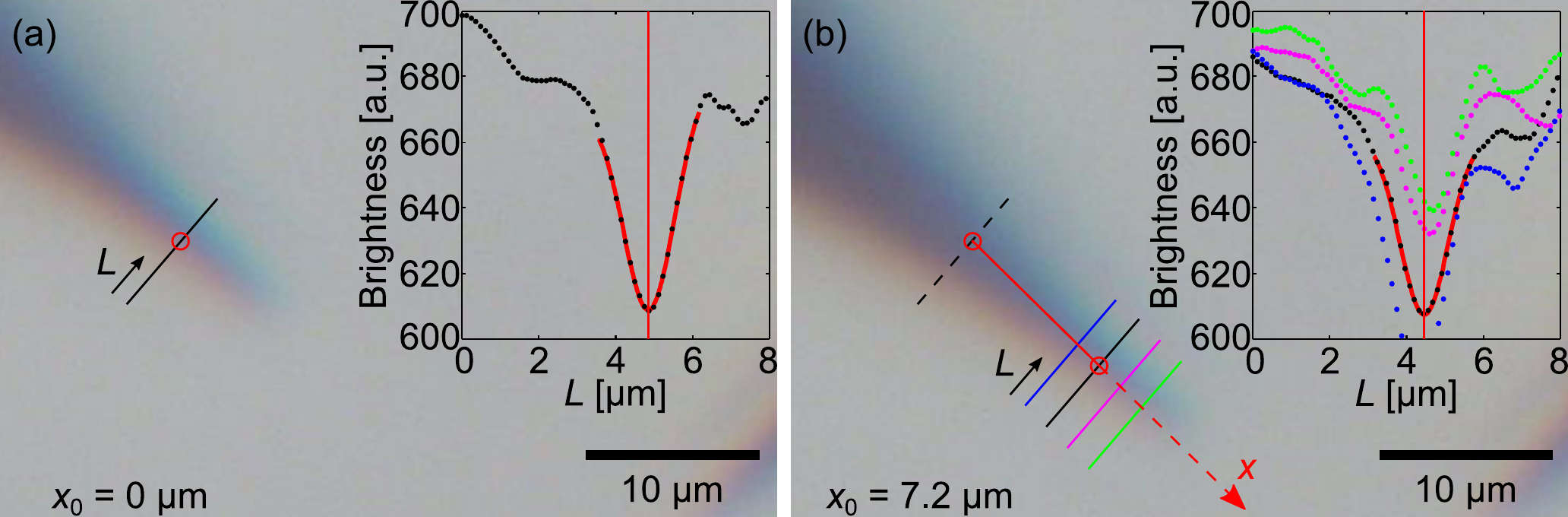}
\caption{Identification of the $x$-axis of the tip movement caused by the piezoelectric actuator. As the tip is moved slowly along the $x$-direction of the scanning system, a series of microscope images is recorded.
\textbf{(a)} In the first micrograph, the brightness values of the image are evaluated along a profile line that is drawn across the tip projection. As shown in the inset, the brightness minimum along the line profile is determined as a reference point for the tip position.
\textbf{(b)} In another micrograph, multiple line profiles that are strictly parallel to the one in the first image are evaluated across the tip projection. The same reference point as defined in the first micrograph is found by identifying the profile line minimum with the same brightness value, as demonstrated in the inset. The $x$-axis of the piezo actuation in the microscope perspective is determined as the connecting line between the recognized reference points in the images (indicated by open circles).}
\label{fig3}
\end{figure*}

A first approximation of distances read from the overview scan is given by the known calibrated linear (low voltage) piezo constant of the corresponding scanner\cite{AFM_book_vo}, which relates the applied voltage to the piezo element from the starting conditions of the overview scan to a nominal lateral displacement $x_0$ and $y_0$ in the surface plane. This constitutes the nominal coordinate system indicated in Fig.~\ref{fig2}~(c). However, as can already be seen from distortions at large $x_0$ and $y_0$ in the large scan image, the actual displacement $x$ and $y$ of the scanning tip follows a non-linear relation to the applied piezo voltage.
To determine this non-linear relation between actual tip displacement $x, y$ and the nominal tip displacement $x_0, y_0$, a quantitative evaluation of the tip displacement as observed by the optical microscope is carried out in the following. 

The key idea is that the tip positions extracted from the optical microscope are real distances, not subject to all the piezo-related distortions. In the following we describe how tip distances can be measured with a precision of $\sim20\,$nm for a typical tip spacing of $1\,\mu$m in spite of the relatively low resolution of the optical microscope of about $\sim 2\,\mathrm{\mu m}$. 
For this purpose, the tip that is intended for recording the overview scan in the experiment is moved along the $x$-axis with a speed corresponding to the scanning motion (e.g. $\sim 30\,$nm/min). By evaluating the relative tip movement in the optical microscope, it is possible to find a relation between the nominal displacement $x_0$ (applied piezo voltage times the low voltage piezo constant) and the actual displacement $x$ (measured with the optical microscope), i.e., a calibration of the movement of the scanning tip, for a specific movement protocol. Constituted by the starting voltage, the voltage range and the speed of voltage application, the movement protocol is chosen to emulate the isolated $x$-motion of the scanning process during acquisition of the overview scan.
Optical micrographs that are taken at different stages of displacement during the tip motion are converted into a matrix containing brightness values that can be used for quantitative analysis.
It is noted that due to piezoelectric hysteresis, the actual tip displacement as a response to an applied piezo voltage is not only non-linear, but also dependent on the initially applied voltage prior to the displacement.
To prevent the influence of hysteresis on the distances determined from the overview scan image and to be able to treat the relation of actual displacement to nominal displacement as reproducible, it is necessary that the overview scan in the experiment is performed under the same conditions as the calibration. Therefore, it is crucial to perform the overview scan with the scanning tip after several hours of rest and strictly according to the same protocol as in the calibration regarding range and speed of the applied piezo voltage. The use of a consistent movement protocol allows to neglect hysteresis effects and ensures that the tip displacement during the overview scan in the experiment follows the same non-linear behavior that is observed in the calibration procedure.

Before the actual displacement can be inferred from the tip movement in optical microscope images, it is important to first identify the orientation of the $x$-axis of the scanning system in the optical microscope image. To do this, the brightness of the first microscope image during $x$-motion is evaluated along a profile line that is drawn across the projection of the tip, as depicted in Fig.~\ref{fig3}~(a). The position of minimum brightness along the profile line is determined as a distinct reference point for the lateral tip position. In the other microscope images that are taken during the movement protocol, the same reference point can be found by sampling multiple profile lines that are parallel to the one in the previous image across the tip projection and finding the corresponding minimum with the same brightness. This is illustrated in Fig.~\ref{fig3}~(b). The connecting line between the minima in the microscope images defines the $x$-axis of the piezo movement.

\begin{figure*}[t]
\includegraphics[width=\linewidth]{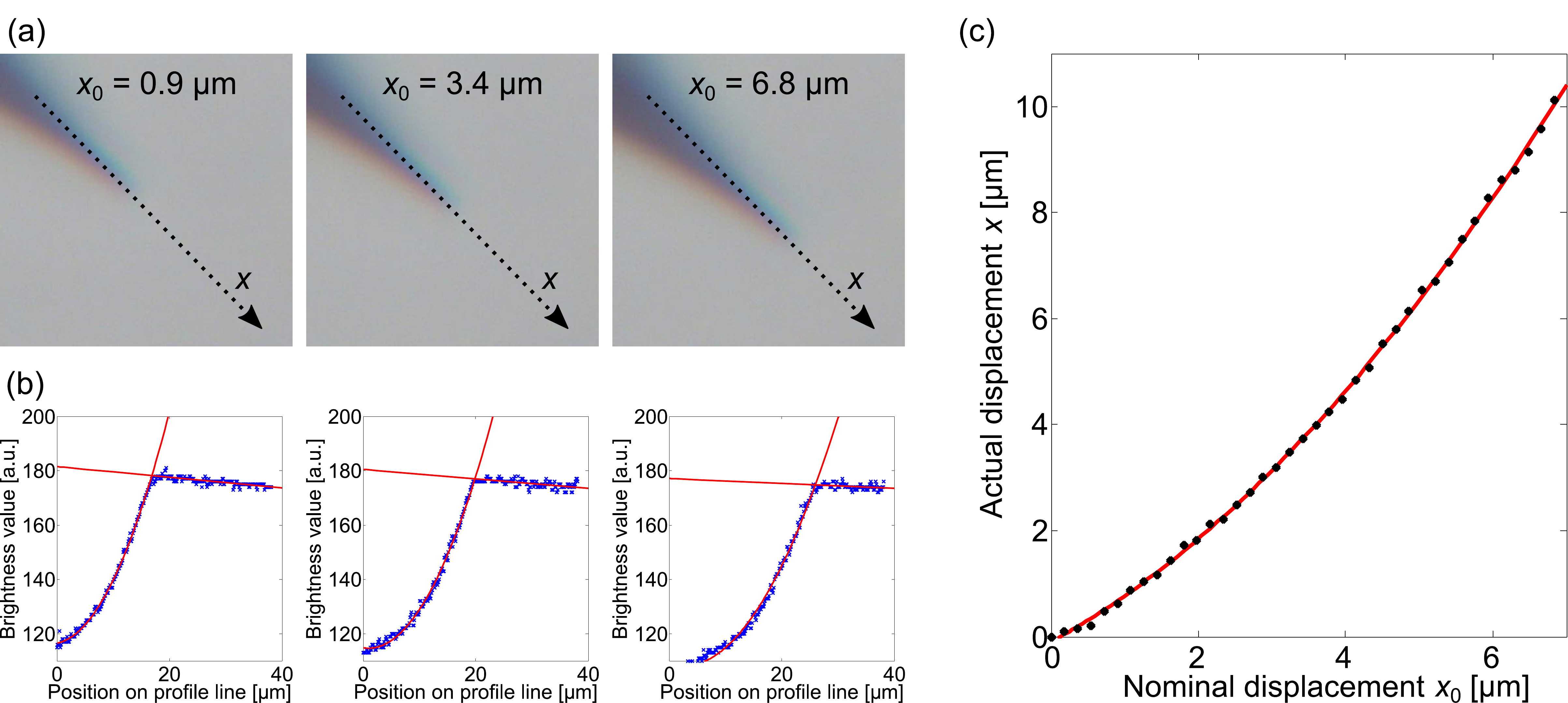}
\caption{Calibration of the non-linear piezo response by quantitative analysis of optical microscope images.
\textbf{(a),(b)} While being moved along the $x$-direction according to a defined protocol, optical microscope images are taken of the tip at different nominal displacement $x_0$. The brightness values obtained from the micrographs along the identified $x$-axis are used to determine the relative movement of the tip projection from one image to another by fitting an empiric model function. The moved distance along the $x$-axis with respect to the the first image corresponding to starting position $x_0 = 0$ is identified as the actual displacement.
\textbf{(c)} The quantitatively obtained values of actual displacement $x$ follow a characteristic non-linear relation to the nominal displacement $x_0$. The actual displacement is modeled heuristically by a second order polynomial function. The obtained set of model parameters is strictly associated to the used movement protocol.}
\label{fig4}
\end{figure*}

With the orientation of the $x$-axis being known, the actual displacement $x$ of the tip can be determined as a function of nominal displacement $x_0$ from the same microscope images as above. For this purpose, the image brightness is evaluated using a profile line through the tip projection \textit{along} the identified $x$-axis in all micrographs, as shown in Fig.~\ref{fig4}~(a) and~(b). The brightness along the profile line consists of an increasing brightness towards the end of the tip, fitted by an empiric model function to the line profile, as well as an approximately constant brightness level beyond the end of the tip, fitted by a straight line. The crossing points of the two red lines in Fig.~\ref{fig4}~(b) correspond to the relative tip displacements with respect to the starting point $x_0 = 0$.
This method of determining the displacement exploits the fact that while the optical microscope does not provide a high resolution to identify the absolute position of a tip, relative changes of the tip positions between the images as obtained from the crossing points of the model functions can be determined with a precision of about 60\,nm.
Another conceivable method of evaluating the actual displacement $x$, instead of the crossing points in Fig.~\ref{fig4}~(b), is the determination of the weighted average location of the distribution of brightness in a small area around the tip. This two-dimensional approach combines the identification of the movement axis in Fig.~\ref{fig3} and the determination of the crossing points in Fig.~\ref{fig4} and as such is essentially equivalent to the method presented in this section.

The actual tip displacement $x$, inferred from this quantitative method, is presented in Fig.~\ref{fig4}~(c) as a function of nominal tip displacement $x_0$. 
The acquired data set is modeled heuristically by a second order polynomial function
\begin{equation}
x(x_0) = a \cdot x_0^2 + b \cdot x_0	,
\label{eq:displacement}
\end{equation}
which is featured as a red curve in Fig.~\ref{fig4}~(c).
Using this model, any nominal tip position $x_0$ in the overview scan in Fig.~\ref{fig2}~(c) can be related to its actual displacement measured from the starting point at the top of the scan.
Distances between tips in $x$-direction can therefore be determined as
\begin{align}
x_i - x_j
&= a \cdot (x_{0 i}^2 - x_{0 j}^2) + b \cdot (x_{0 i} - x_{0 j})\\
\Leftrightarrow \qquad \ \ s
&= a \cdot s_0 (x_{0 i} + x_{0 j}) + b \cdot s_0	,
\end{align}
with $s_0$ and $s$ denoting the nominal and actual tip distance between a pair of tips indexed by $i$ and $j$. It is emphasized that this model is understood to reproduce actual tip displacement in $x$-direction during the overview scan only for the specific movement protocol (speed of motion) it is calibrated to.
Further, it is stressed that the actual navigation of the tips into target positions -- after the overview scan is recorded -- is carried out by using topographic features as beacons, as demonstrated in Fig.~\ref{fig2}~(d)-(f). Due to the hysteretic behavior of the piezoelectric actuators, the distance covered by tips generally depends on their previous action. Therefore, it is not advisable to rely on the nominal tip displacements according to Eq.~\ref{eq:displacement} as a means of navigation.

\begin{figure*}[t]
\includegraphics[width=\linewidth]{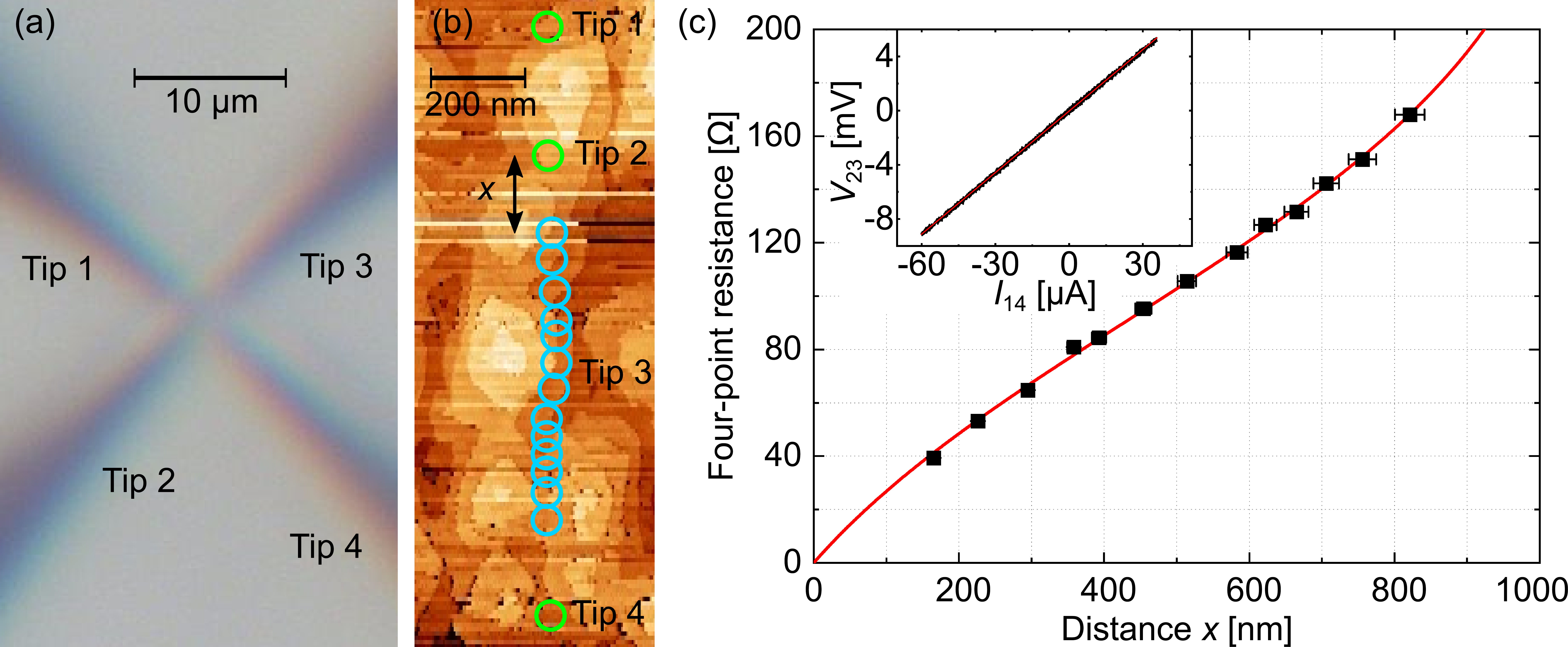}
\caption{Distance-dependent four-point resistance measurement of a 10\,nm thick  BiSbTe$_3$ film on the nanometer scale.
\textbf{(a)} Final tip configuration resulting from positioning via STM scans as viewed by the optical microscope. With all four tips placed on the surface within a linear section of just $1.3\,\mu$m of length, details of the tip arrangement are completely indiscernible in the optical view.
\textbf{(b)} Section of the overview STM scan that is used for the alignment of the tips. The positions of the static tips in the electrical measurements are marked by green circles, while the position of the mobile tip is indicated in cyan. In the experiment, a current is injected between the outer two tips, with the inner two tips probing the resulting potential difference.
\textbf{(c)} Measured four-point resistance as a function of the distance $x$ between the two voltage-probing tips. A corresponding representative $I$-$V$ curve is presented in the inset. The acquired data is explained by a 2D conductivity model represented by the red curve. The conductivity of the sample is determined to $(2.76 \pm 0.05)\,$mS.}
\label{fig5}
\end{figure*}

The uncertainty of tip distances $\sigma_s$ determined in this manner is influenced by the uncertainties of the model parameters $\sigma_a$ and $\sigma_b$ from the fit and is obtained from the propagation of errors as
\begin{equation}
\frac{\sigma_s}{s} = \frac{\sqrt{\left( s_0 + 2x_{0j} \right)^2 \sigma_a^2 + \sigma_b^2}}{a \left( s_0 + 2x_{0j} \right) + b}	.
\end{equation}
For a set of parameters $a = (8.2 \pm 0.3)\cdot 10^{-5}\,\mathrm{nm}^{-1}$ and $b = (0.77 \pm 0.02)$ obtained for the exemplary movement protocol presented in Fig.~\ref{fig4}~(c), the relative uncertainty of determined distances amounts to $\sim 2.5\%$. For tip configurations with typical nominal inter-tip distances of $s_0 \approx 1\,\mu$m, this results in an uncertainty of $\sigma_s \approx 20\,$nm.
The entire displacement calibration procedure relies on the principle that contrary to absolute positions, relative changes in the tip position between several images can be determined confidently by quantitative analysis from optical micrographs.

In principle, the calibration of a defined movement protocol could also be done by applying a pattern recognition algorithm to the series of micrographs as a more sophisticated way of determining tip displacement.
It is further noted that it is only necessary to characterize the non-linearity of the piezo actuator corresponding to the tip which is performing the large overview scan. Distortions in the smaller scans acquired by the other tips are not relevant, since the determination of tip positions is made via the identification of topographic features in the overview scan.
In typical multi-probe transport experiments, linear tip configurations are realized. In that case, the symmetry axis of the tip can be chosen to correspond with the $x$-axis of the overview scan when positioning the tips on the sample surface and inter-tip distances can be inferred from the calibrated scan image. 
If required, e.g., if measurements with non-linear tip configurations are planned, an analogous calibration of the $y$-motion of the scanning tip as a function of nominal displacement $y_0$ can be done using the same principle. For this purpose, the tip intended for recording the overview scan is moved along the $y$-direction to emulate scanning motion and the corresponding axis is identified in the same manner as presented in Fig.~\ref{fig3}. Subsequently, optical micrographs that are acquired during the motion are used to evaluate the image brightness along a profile line in $y$-direction, i.e., \textit{across} the tip shape. Similar to the insets of Fig.~\ref{fig3}, a Gauss curve can be modelled to the brightness data following the profile line. The coordinate of the related brightness minimum from each micrograph corresponds to the relative tip displacement with respect to the starting point $y_0 = 0$. Consequently, the nominal displacement $y$ is obtained as a function of $y_0$ for the specific movement protocol and as such, the overview scan corresponding to the protocol is calibrated in two dimensions.

\section*{Other sources of image distortion}

Apart from the non-linear piezo response, a displacement caused by the piezoelectric effect is subject to creep, thermal drift, and hysteresis effects \cite{AFM_book_vo}.
These sources of error for distance determination from the scan image are evaluated in the following.
Piezoelectric creep is an effect that only persists for a relatively short time after applying a voltage to the scanning piezo. The impact of the initial creep on the recorded overview scan can be circumvented by waiting for a sufficient amount of time before starting the scan.
Also, even if an initial creep effect is present at the start of the overview scan image, its influence on the determined distances can be easily minimized by realizing a measurement configuration in the center of the overview scan, where the effect has already faded out.

Thermal drift in the xy-plane as a further potential source of image distortion is observed to be constant at $< 0.3\,$nm/min in the experiment \cite{Cherepanov2012} for the relevant time scale of the overview scan.
With an average scanning speed along the slow axis of $\sim 30\,$nm/min, the resulting distortion of the scan image arising from thermal drift amounts to less than $1\%$. This constant drift effect enters the actual tip displacement $x$ systematically and is captured by the calibration procedure as well. While a constant systematic thermal drift is accounted for in the calibration, possible variations of the drift speed constitute an additional statistical error. Since these deviations from the average drift speed are (much) smaller than the drift itself, thermal drift as a source of error for the determination of inter-tip distances is therefore regarded as insignificant.

Finally, the influence of hysteresis on the distances determined from the scan image can be prevented by strictly following the same movement protocol, in both, the calibration procedure and the acquisition in the overview scan in the experiment, as explained in the previous section.

\section*{Example measurement}

An exemplary distance-dependent four-point resistance measurement realized by the demonstrated positioning technique is presented in this section.
Using the presented positioning technique, four-point resistance measurements can be performed in tip configurations with much smaller tip distances than the ones which can be realized with position control by optical microscopy and sometimes even by SEM. By using piezoelectric control to navigate tips across the mapped surface in tunneling contact and affirm their positions by performing small local scans, minimum distances between the tips in the experiment are only limited by their radii. 

Figure~\ref{fig5} presents results of an exemplary distance-dependent four-point resistance measurement on the surface of a template BiSbTe$_3$ sample.
The optical microscope view of the linear tip arrangement with all four tips aligned within $1.3\,\mu$m is seen in Fig.~\ref{fig5}~(a), after all tips have been moved to their final measurement positions. Evidently, the ends of the tips are indiscernible in the image for the final tip configuration.
The positions of the tip in a section of the overview scan are indicated by colored circles in Fig.~\ref{fig5}~(b).

Within the scope of the corresponding electrical measurements, all tips are in contact with the surface and a current is induced between the outer two tips (tips 1 and 4) of the linear arrangement while simultaneously probing the potential difference between the inner two tips (tips 2 and 3) separated by the distance $x$. The four-point resistance is subsequently obtained from a linear fit to the $I$-$V$ curve, as presented exemplarily in the inset of Fig.~\ref{fig5} (c). In this manner, the resistance is acquired at variable $x$ by moving tip 3 along the configuration symmetry line in tunneling contact between each electrical measurement. The measured resistance of the sample as a function of inter-tip distance $x$ is presented in Fig.~\ref{fig5} (c). It is possible to probe the local resistance at more than $10$ points for the moved voltage probing tip (tip 3). As described in the previous section, the distance $x$, as well as the distances $s_{12}$ and $s_{24}$ between the static tips, are determined from the overview scan image.
The acquired data is modeled by applying the general expression for the four-point resistance of a two-dimensional system to the geometry of the tip configuration, which results in \cite{Leis2020}
\begin{equation}
R_\mathrm{2D}(x) = \frac{1}{2 \pi \sigma_\mathrm{2D}} \left[ \ln\left( \frac{s_{24}}{s_{12}} \right) - \ln\left( \frac{s_{24} - x}{s_{12} + x} \right) \right]	.
\end{equation}
From a fit of the model to the data, the two-dimensional conductivity of the sample is determined to $(2.76 \pm 0.05)\,$mS, which coincides well with measurements at macroscopic distances on the same sample \cite{Leis2020}.

\section*{Conclusion}

In conclusion, we find that using topography scans performed with each of the individual tips of a four-tip STM to monitor tip positions enables four-point measurement configurations on the nanometer scale without the need for an SEM. 
The use of a calibrated STM scan as a reference map for the navigation of tips opens a large range of possible tip configurations for transport measurements with tip-tip distances only limited by the tip radii. The corresponding inter-tip distances are determined from the topography scan, with a precision of $20\,$nm at a typical tip spacing of $\sim 1\,\mu$m. This method also prevents systematic errors in the determination of tip-tip distances due to bent tips (c.f. Fig.~\ref{fig1}~(b)). Moreover, the use of an optical microscope and STM scanning prevents any contamination issues due to an electron beam or any impact of the electron beam on the electric properties of the sample, as it can occur due to the use of SEM imaging.
For future studies based on four-tip STM, the demonstrated tip positioning method can be used to contact delicate nanostructures precisely, allowing for the measurement of effects in charge transport that are restricted to small dimensions.

\begin{acknowledgments}
The authors gratefully thank Michael Schleenvoigt and Gregor Mussler for providing the template sample used for the demonstration of the positioning method.
This work is funded by the Deutsche Forschungsgemeinschaft (DFG, German Research Foundation) under Germany's Excellence Strategy - Cluster of Excellence Matter and Light for Quantum Computing (ML4Q) EXC 2004/1 - 390534769. F.S.T. acknowledges support of the Deutsche Forschungsgemeinschaft through the SFB 1083, project A12.
\end{acknowledgments}

%

\end{document}